# UM BANCO DE DADOS DE EMPREGOS FORMAIS GEORREFERENCIADOS EM CIDADES BRASILEIRAS

**André Borgato Morelli**[*]
**André de Carvalho Fiedler**
**André Luiz Cunha**
Universidade de São Paulo
Escola de Engenharia de São Carlos

**ABSTRACT**

Currently, transport planning has changed its paradigm from projects oriented to guarantee service levels to projects oriented to guarantee accessibility to opportunities. In this context, a number of studies and tools aimed at calculating accessibility are being made available, however these tools depend on job location data that are not always easily accessible. Thus, this work proposes the creation of a database with the locations of formal jobs in Brazilian cities. The method uses the RAIS jobs database and the CNEFE street faces database to infer the location of jobs in urban regions from the zip code and the number of non-residential addresses on street faces. As a result, jobs can be located more accurately in large and medium-sized cities and approximately in single zip code cities. Finally, the databases are made available openly so that researchers and planning professionals can easily apply accessibility analyzes throughout the national territory.

**RESUMO**

Atualmente, o planejamento de transportes tem mudado seu paradigma de projetos orientados a garantir níveis de serviço para projetos orientados a garantir acessibilidade a oportunidades. Nesse contexto, um sem número de estudos e ferramentas voltadas ao cálculo de acessibilidade estão sendo disponibilizadas, contudo essas ferramentas dependem de dados de localização de empregos que nem sempre são de fácil acesso. Assim, este trabalho propõe a criação e disponibilização de um banco de dados com as localizações de empregos formais nas cidades brasileiras. O método toma os bancos de dados de empregos da RAIS e de faces de logradouros do CNEFE para inferir a localização de empregos em regiões urbanas a partir do CEP e do número de endereços não residenciais em faces de logradouro. Como resultado, é possível localizar empregos com maior precisão em cidades grandes e médias e de forma aproximada em cidades de CEP único. Por fim, os bancos de dados são disponibilizados de forma aberta para que pesquisadores e profissionais de planejamento possam aplicar análises de acessibilidade de maneira facilitada em todo território nacional.

---

[*] Autor correspondente: andre.morelli@usp.br

## 1. INTRODUÇÃO

Por muito tempo, o planejamento de sistemas de transporte urbano foi regido integralmente por normas que se baseiam em parâmetros de mobilidade como conforto do usuário no trajeto, velocidade de cruzeiro, atrasos em interseções e níveis de serviço em geral (TRB, 2022, 2016, 2010). Existe, contudo, uma forte tendência na literatura atual em direção à modelagem baseada em acessibilidade (Barboza *et al.*, 2021; Boisjoly *et al.*, 2020; Braga *et al.*, 2020; Carneiro *et al.*, 2019; Geurs e Ritsema van Eck, 2001; Geurs e van Wee, 2004; Hansen, 1959; Levine *et al.*, 2012; Pritchard *et al.*, 2019; Sarlas *et al.*, 2020). A mudança de paradigma que essa tendência causa é na consideração do papel efetivo que o uso do solo urbano tem nos movimentos dos usuários. Isso se dá porque, enquanto as características associadas à mobilidade também são importantes para garantir acessibilidade, o tempo/distância de trajeto também tem papel fundamental na análise.

Na análise tradicional focada em mobilidade, um trajeto entre casa e trabalho pode ter uma duração significativa desde que seja realizada em nível de serviço alto, o que leva o modelo tradicional de planejamento a favorecer o espraiamento urbano ao afastar usuários de seus destinos no esforço de reduzir a concentração de usuários em poucos elementos de infraestrutura viária (Carneiro *et al.*, 2019; Levine *et al.*, 2012; Nadalin e Igliori, 2015, 2010). Análises voltadas a acessibilidade, por outro lado, por mais que reconheçam a importância da mobilidade, tendem a dar importância à proximidade entre origens e destinos, sendo mais tolerantes com velocidades baixas e até mesmo congestionamentos desde que os tempos de viagem sejam relativamente curtos, favorecendo a densidade populacional e proximidade entre usuários e oportunidades (Levine *et al.*, 2019, 2012). Isso significa que, apesar de as duas abordagens compartilharem em grande parte as ferramentas de análise, o enfoque em uma ou em outra leva a resultados significativamente diferentes.

Nesse contexto, análises de acessibilidade exploram o problema do transporte urbano de maneira mais completa, contudo, os avanços para a popularização de análises de acessibilidade na prática têm sido lentos (Handy, 2020). Em grande parte, isso se dá porque análises de acessibilidade necessitam de um conjunto de dados mais extenso para serem aplicadas, além de requererem esforços computacional e técnico significativos para explorar soluções.

Ferramentas computacionais têm sido extensivamente desenvolvidas para o cálculo de acessibilidade, como aplicações implementadas tanto na linguagem Python (Blanchard e Waddell, 2017) quanto na linguagem R (Pereira *et al.*, 2021) além de ferramentas para computação de métricas de rede e características de tráfego nesses ambientes (Boeing, 2017; Csardi e Nepusz, 2005; Liu *et al.*, 2021; Waddell *et al.*, 2018). Contudo, para que haja um salto na aplicação de métricas de acessibilidade para guiar políticas públicas no Brasil, é necessário que bancos de dados de fácil acesso sejam disponibilizados para tomadores de decisão.

Nesse contexto, com o intuito de facilitar o planejamento de transportes pela perspectiva da acessibilidade em cidades brasileiras, este trabalho propõe a disponibilização de um banco de dados contendo a localização de empregos formais em cidades brasileiras. O método proposto abrange a inferência da localização dos empregos a partir do código postal da empresa empregadora cadastrado na Relação Anual de Informações Sociais (RAIS) e o número de endereços com fins não residenciais em faces de logradouro. A maior contribuição deste trabalho está na inferência a partir dos endereços não comerciais que possibilita obter uma aproximação da localização dos empregos mesmo em cidades com CEP único, apesar de haver nesses casos prejuízo à qualidade da inferência. Assim, o banco de dados disponibilizado como produto final deste trabalho se divide em duas categorias: uma compreendendo cidades de médio e grande porte e outra compreendendo cidades menores. Essa divisão se deve à maneira como os códigos postais são atribuídos a localizações no Brasil dado que cidades de maior porte tendem a ter um ou mais CEPs por logradouro enquanto cidades pequenas (menores de 50 mil habitantes) possuem CEP único para o município. Isso significa que a qualidade dos dados para o grupo que compreende cidades menores tende a ser inferior à de cidades maiores, contudo espera-se que essa inferência inicial possa server de suporte para análises e projetos iniciais em cidades de pequeno porte.

## 2. MÉTODO

Desde 2014, a tabela da Relação Anual de Informações Sociais (RAIS) contém o campo do CEP das empresas (MTE, 2014), possibilitando a geolocalização de empregos em território nacional.

Contudo, a geolocalização por CEP é um processo impreciso por dois motivos: primeiro, as coordenadas referentes aos CEPs não são disponibilizadas gratuitamente pelos Correios, de forma que a geolocalização fica dependente do banco de dados de outras empresas ou agências; segundo, um CEP em geral está associado a um logradouro que pode se estender por vários quilômetros ou, em casos de cidades de pequeno porte, o CEP está associado ao município todo.

Para lidar com esses problemas, a geolocalização dos empregos foi conduzida com auxílio de um banco de dados secundário, proveniente do Cadastro Nacional de Endereços para Fins Estatísticos (CNEFE) atrelado ao IBGE (CNEFE, 2020). Em 2019 o CNEFE disponibilizou os dados geoespaciais de faces de logradouros em todo território nacional, com arquivos geográficos da divisão de logradouros em faces de tamanho reduzido (tamanho de um quarteirão) com informações como o número de endereços residenciais e número de endereços totais, como ilustrado na Figura 1. O Cadastro também fornece um banco de dados separados por vírgula com a espécie dos endereços associados a cada face de logradouro, dividida em sete categorias: (1) domicílio particular; (2) domicílio coletivo; (3) estabelecimento agropecuário; (4) estabelecimento de ensino; (5) estabelecimento de saúde; (6) estabelecimento de outras finalidades; (7) edificação em construção. Esse banco de dados também fornece o CEP desses endereços.

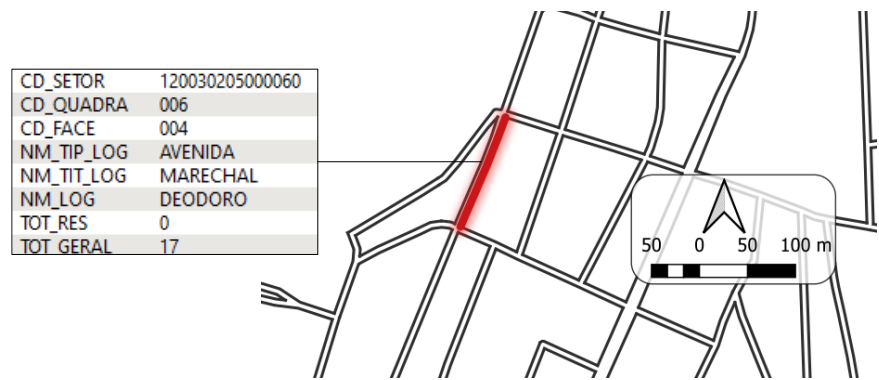

Figura 1: Exemplo de uma face de logradouro.

Com esses dois bancos de dados, é possível conduzir um processo de fusão para a geolocalização do número e espécie de empregos na malha urbana. Para tanto os empregos referentes a uma empresa da tabela RAIS são associados às faces de logradouro do banco de dados do CNEFE que possuem mesmo CEP, que pode ser feito de maneira igualitária em todas as faces com mesmo

CEP, como demonstrado na Figura (b). Contudo, para tentar aumentar a precisão da localização, neste trabalho os empregos são distribuídos de forma ponderada pelo número de endereços comerciais de espécie compatível com a da empresa, como mostrado no diagrama da Figura 2 (c). Esse processo ainda está submetido a imprecisões, já que as categorias do CNEFE são relativamente amplas, com estabelecimentos comerciais figurando na mesma categoria de estabelecimentos industriais (espécie "outras finalidades") contudo é um processo melhor de distribuição de que as alternativas que seriam concentrar todos os empregos no ponto médio do logradouro ou distribuir uniformemente ao longo do logradouro.

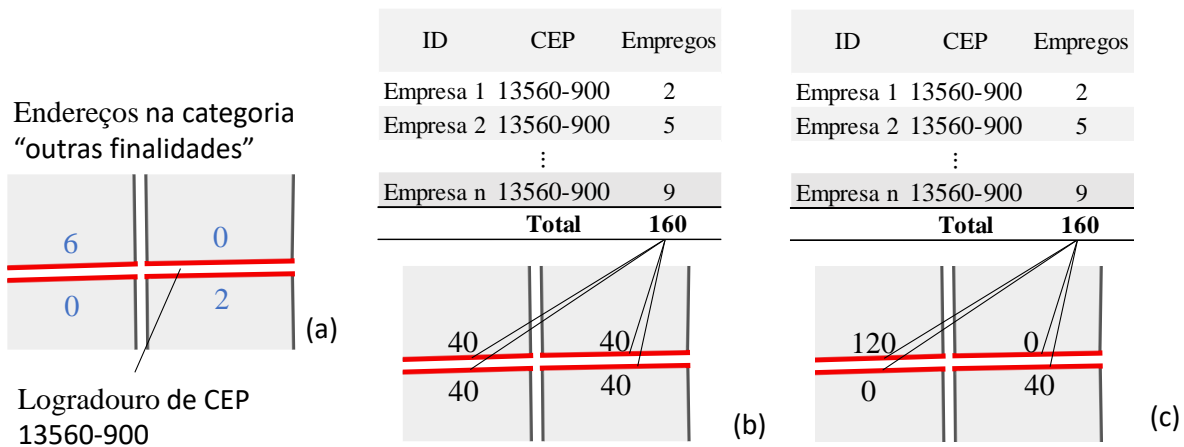

Figura 2: Modo de distribuição de empregos a partir de endereços não residenciais.

No caso de cidades de CEP único, o problema de a espécie "outras finalidades" abranger diversos tipos de empreendimento se intensifica, principalmente no que diz respeito a endereços industriais e comerciais. Endereços industriais em geral são localizados em distritos específicos nas cidades de pequeno porte e, além disso, indústrias tendem a concentrar um grande volume de empregos em poucos endereços. Com a distribuição proposta neste trabalho, os empregos são distribuídos de forma proporcional ao número de endereços e, portanto, existe a tendência a superestimar o número de empregos em regiões de serviços (centros comerciais) e subestimar empregos em regiões industriais de cidades pequenas. Esse problema é praticamente inexistente em cidades de múltiplos CEPs (médias e grandes) já que é possível identificar o distrito industrial pelos CEPs das empresas ali localizadas.

## 3. RESULTADOS

Como resultado, o banco de dados contém as informações sobre empregos nas cidades brasileiras referenciado a partir de um ponto no espaço (latitude/longitude) que representa a face de logradouro em que se acumulam esses empregos. A Tabela 1 contém as cinco primeiras linhas do banco de dados referente à cidade de São Carlos-SP no formato csv (valores separados por vírgula).

Tabela 1: Primeiras cinco linhas do arquivo de valores separados por vírgula da cidade de São Carlos-SP

| cod_face | CEP | non residential | jobs | lon | lat |
|---|---|---|---|---|---|
| 172100005000001000000 | 77001422 | 2 | 8 | -48.3533 | -10.1651 |
| 172100005000002000000 | 77001440 | 1 | 3 | -48.3506 | -10.1691 |
| 172100005000003000000 | 77001390 | 1 | 4 | -48.349 | -10.1667 |
| 172100005000004000000 | 77001392 | 1 | 4 | -48.3533 | -10.164 |
| 172100005000005000000 | 77001566 | 2 | 8 | -48.353 | -10.1622 |

Os dados também são disponibilizados nos formatos shapefile (.shp) e geopackage (.gpkg) mais facilmente carregados em softwares de Sistemas de Informação Geográfica (SIG). A Figura 3 contém a representação geográfica dos pontos com empregos na cidade de São Carlos-SP.

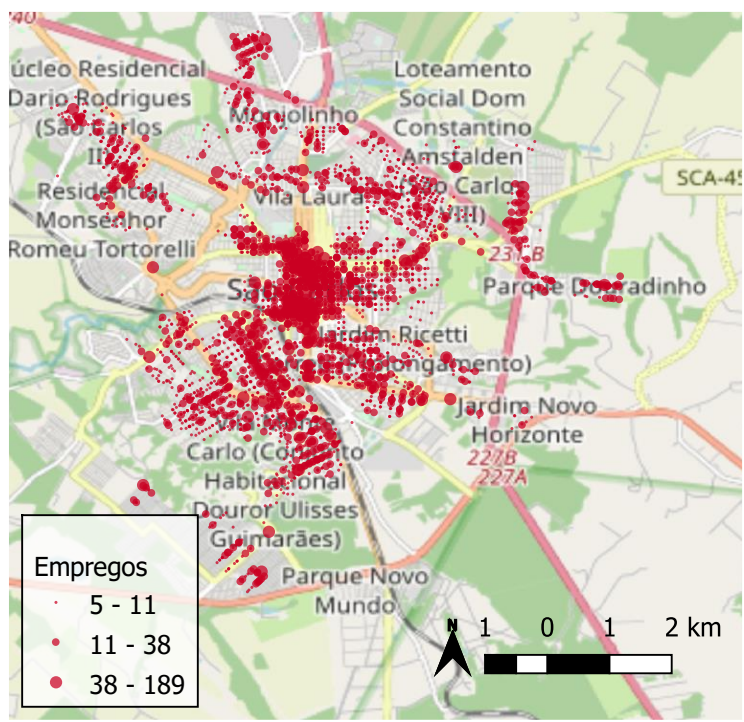

Figura 3: Modo de distribuição de empregos a partir de endereços não residenciais.

## 4. DISPONIBILIDADE DO BANCO DE DADOS

Por hora, os bancos de dados de qualquer cidade brasileira ou conjunto de cidades podem ser acessados sob demanda ao autor correspondente sem limitações com relação ao uso do material desde que este trabalho seja devidamente citado. Ressalta-se que os dados estão disponíveis apenas a partir do ano de 2014, primeiro ano em que a RAIS disponibilizou os CEPs das empresas cadastradas. Espera-se que em breve os dados possam ser disponibilizados abertamente em uma plataforma dedicada assim que os autores encontrem um domínio acessível.

## 5. CONCLUSÃO

Neste trabalho foi apresentado um método para a catalogação e georreferenciamento de empregos em cidades brasileiras a partir do Cadastro Nacional de Endereços para Fins Estatísticos (CNEFE) e da Relação Anual de Informações Sociais (RAIS). O método possui o diferencial de ser capaz de inferir a posição dos empregos em um determinado logradouro a partir do número de endereços não residenciais associados a cada face de logradouro, o que possibilita localizar empregos com maior precisão e possibilita que o processo seja executado mesmo em cidades de CEP único.

No futuro os bancos de dados serão disponibilizados de maneira aberta em uma plataforma dedicada, contudo até lá os dados podem ser obtidos a partir de consulta ao autor correspondente.

## REFERÊNCIAS